# Spectra of pyroelectric X-ray generator

Nagaychenko V.I., Sanin V.M., Yegorov A.M., Shchagin A.V.

Kharkov Institute of Physics and Technology,

Kharkov 61108, Ukraine

**Abstract**

The construction of X-ray generator based on a pyroelectric crystal $LiNbO_3$ is described. Some properties of radiation spectra from the X-ray generator are presented. Measurements of the spectra were performed at heating and cooling of the crystal with copper and chromium targets. The maximum energy in the X-ray spectrum versus crystal temperature is presented.

**Introduction**

The pyroelectric crystals are known to have an amazing capability for generating electron beams, as their temperature changes [1-8]. The electron beam energy can attain 170 keV [7], depending on the type of crystal, its thickness, heating/cooling conditions. The electron beam can even be self-focused [8]. The electric fields generated by pyrocrystals can be as high as $10^6$ V/cm [9]. As it has been recently demonstrated, this phenomenon can be used for creating a compact low-power X-ray generator [6,9,10]. The basic peculiarities of this generator are the absence of outer high-voltage power source, safety, small size [6,9,10]. Here, we present some properties of X-ray spectra from pyroelectric X-ray generator with copper and chromium foils.

**The X-ray generator**

Pyroelectric X-ray generator was created at Kharkov Institute of Physics and Technology. The scheme of the generator is similar to one described in Refs. [6, 9, 10]. It is shown in Fig. 1.

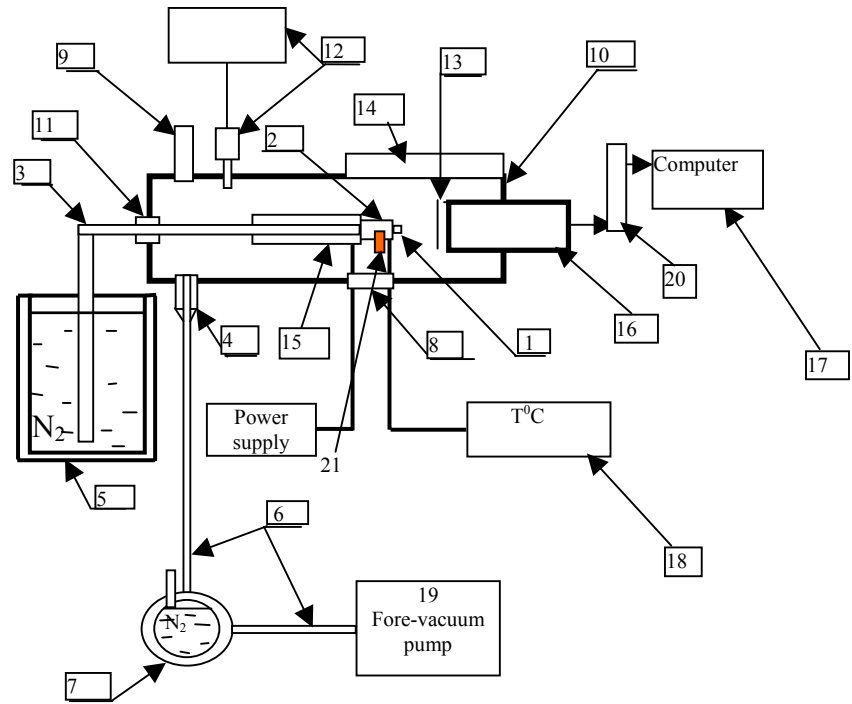

Fig. 1. A scheme of the pyroelectric X-ray generator.
1 - $LiNbO_3$ crystal, 2 - brass base, 3 - heating/cooling line, 4 – vacuum valve, 5 - Dewar vessel with liquid nitrogen, 6 - vacuum line, 7 - liquid-nitrogen trap, 8 - vacuum connector, 9 –variable leak valve, 10 - vacuum chamber of the generator, 11 - teflon vacuum obturator of the cooling line, 12 – gas pressure measuring device, 13 - metal foil, 14 - transparent glass window, 15 - resistor-heater, 16 - Si(Li) X-ray detector, 17 - computer, 18 - temperature-measuring device, 19-fore-vacuum pump, 20 - pulse-height analyzer, 21 – thermocouple.

The generator uses a pyroelectric crystal $LiNbO_3$ of size 3.7x4.3x2.5 mm$^3$, which is subjected to cyclic variations of temperature (heating/cooling) in the temperature range from - 50$^\circ$C to + 120$^\circ$C. The Z – axis of the crystal is along the axis of the generator. A vitrified wire resistor was used for heating, while liquid nitrogen was used for cooling. The cooling line was fabricated from copper rod, and its one end was closely connected with the crystal and the heater, and the other end was immersed in liquid nitrogen. The duration of heating was specified by the resistor current value. The air pressure in the chamber of the generator can be regulated with the help of a variable leak valve and a vacuum pump. The optimum pressure ranges was found between 10 and 50 mTorr. The grounded metal foil serve as a target for generation of X-rays. The spacing between the foil and the crystal surface can be varied from 0 to 15 mm. The X-ray spectra are measured by a semiconductor Si(Li) X-ray detector having a resolution of 250 eV (full width at half height). The brass base and cooling line and target was grounded.

### X-ray spectra

The crystal surface on the side of the detector should be charged positively at heating and negatively at cooling. So, at cooling, the electrons can go from the crystal to the target. The target, being bombarded with electrons emits the characteristic X-ray radiation and bremsstrahlung. X-ray radiation spectra measured at cooling with Cu and Cr targets are shown in Figs. 2,4 respectively.

At heating, the crystal polarization is reversed and surface, which directed to target, should be charged positively. In this case, characteristic X-ray radiation and bremsstrahlung goes from the crystal. This radiation passes trough a thin foil and reaches the detector. At crystal heating, the radiation spectra comprises the characteristic K-lines of both crystal and target elements (Figs. 3,5). A part of radiation can go from the chamber walls, as the body of the vacuum chamber is made with brass. The registered radiation intensity corresponding to cooling is higher than the one corresponding to heating.

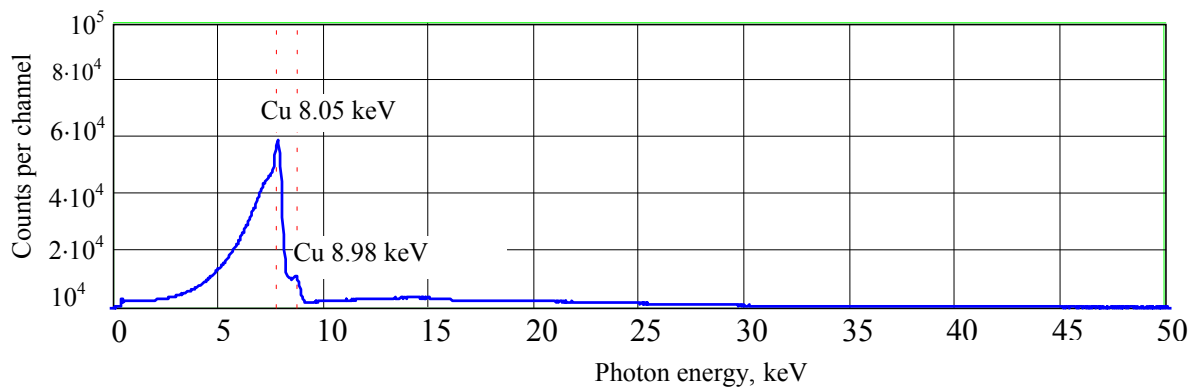

Fig. 2. X-ray radiation spectrum measured with Cu target thickness 20 μm at $LiNbO_3$ crystal cooling. The value of one channel is 0.036 keV. The distance between the crystal and the copper foil is 7.5 mm. The total number of counts in the spectrum is 8.391·10$^6$ for 16 minutes. The residual air pressure in the vacuum chamber is measured to be 25 mTorr.

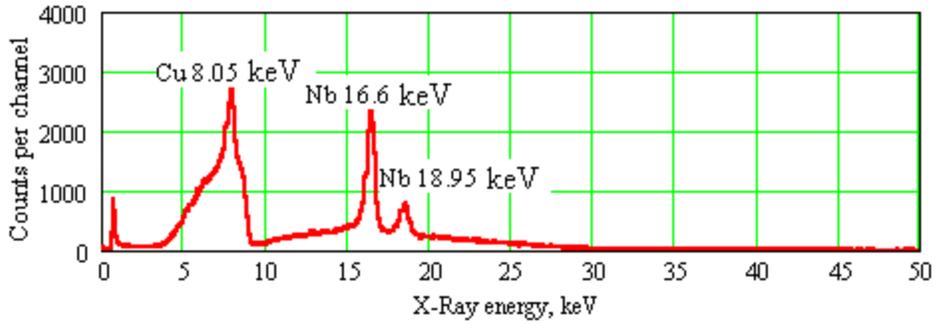

Fig. 3. X-ray radiation spectrum measured with Cu foil and crystal elements during LiNbO$_3$ crystal heating. The distance between the crystal and the copper foil is 7.5 mm. The total number of counts in the spectrum is $4.975 \cdot 10^5$ for 8 minutes. The residual air pressure in the vacuum chamber, which accommodates the crystal, is measured to be 25 mTorr.

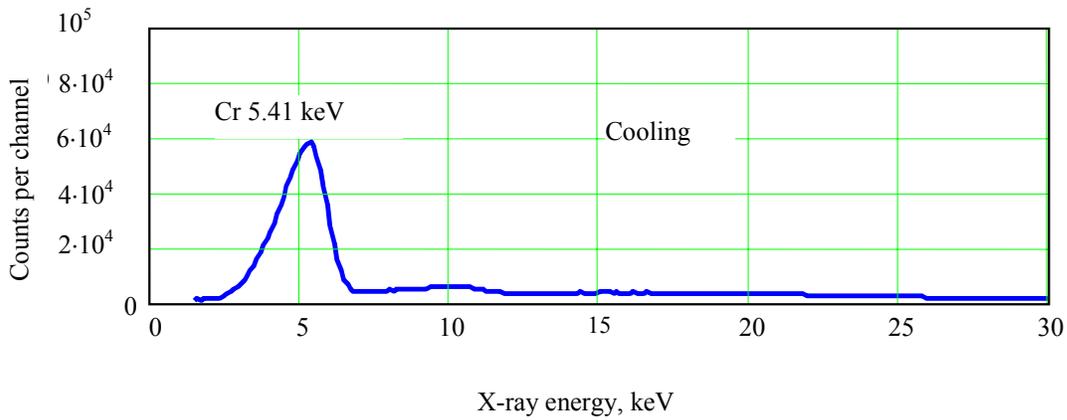

Fig. 4. X-ray radiation spectrum measured with Cr target thickness 18 μm at LiNbO$_3$ crystal cooling for one minute.

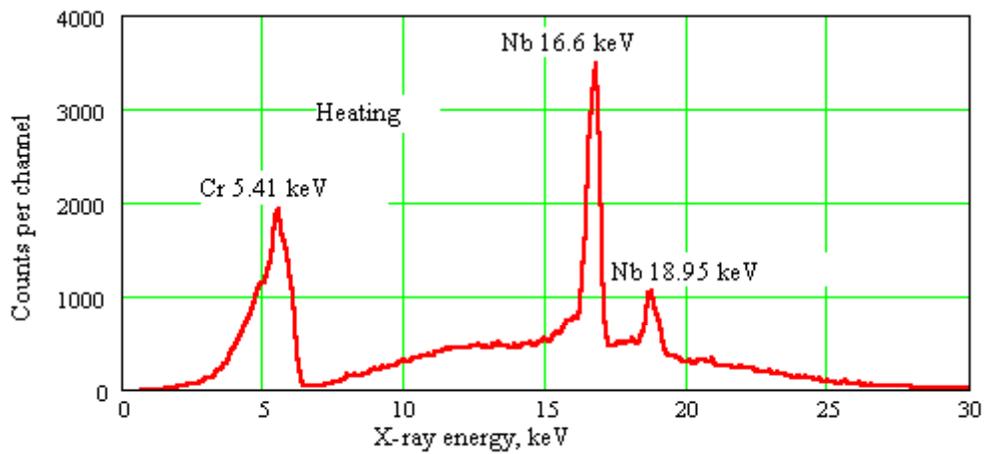

Fig. 5. X-ray radiation spectrum measured with Cr target thickness 18 μm at LiNbO$_3$ crystal heating for one minute.

**The maximum energy of x-rays**

Experiments were made to measure the x-ray spectra at dynamic conditions. During a thermal cycle, the several spectra were measured for one minute each.

Eight spectra were measured at LiNbO$_3$ heating from - 50°C up to +125°C. Fig. 6 shows a typical dependence of measured maximum energy of X-rays in spectra on the temperature. In this experiment, the crystal of size 3.7x4.3x2.5 mm$^3$ was used. It is seen from the figure that as the crystal was heated, the maximum energy increased up to ~28 keV (20°C), and then fell off to 17 keV (62°C) with a further rise up to ~ 30 keV (120°C). After heating was ceased, the intensity and maximum energy of the x-rays went to zero.

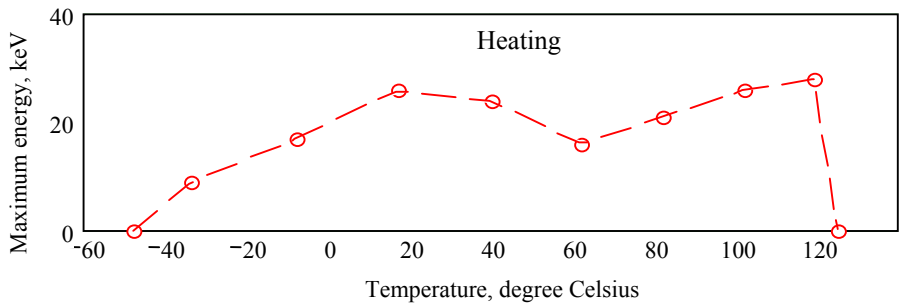

Fig. 6. Maximum energy of X-rays versus crystal temperature at heating conditions.

Similar measurements were also conducted on cooling the crystal. A typical curve for the measured maximum energy of radiation versus crystal temperature at cooling conditions is shown in Fig. 7. It increases, as the crystal is cooled, and reaches its maximum of ~38 keV at room temperature ( ~20°C), then the x-ray maximum energy drops to 0 keV (- 40°C).

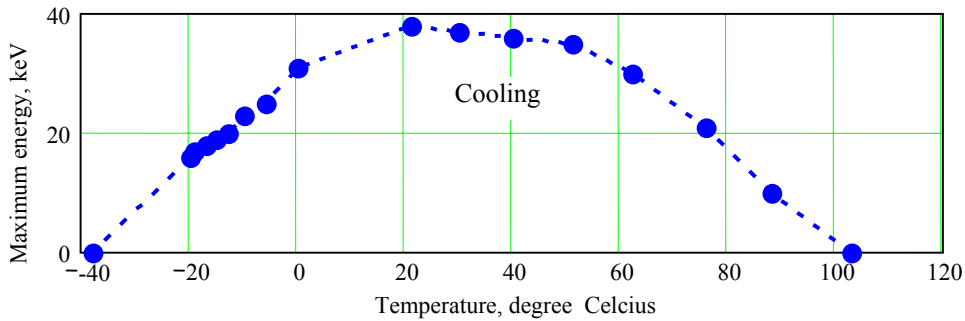

Fig. 7. Maximum energy of X-rays versus crystal temperature at cooling conditions.

**Conclusion**

The pyroelectric X-ray generator was created in KIPT and its operation was demonstrated. Results of our first experiments are mainly in agreement with the results of earlier pioneer works [6, 8-10].

Continuous part of X-ray spectra in Figs. 2-4 should be due to bremsstrahlung of accelerated electrons in a target and crystal. Thus, the maximum energy of X-rays should be close to maximum energy of these electrons. Therefore, Figs. 6,7 show values close to maximum energy of electrons in the generator.

## Acknowledgements

We are thankful to S.M. Shafroth and J.D. Brownridge for encourage to start works in this field and J.D. Brownridge for gift of pyrocrystals for our first experiments, and to Crystal Technology Inc. what supplied us with pyrocrystal samples. The work was performed on STCU 1911 project.